# SIMPLIFYING AND IMPROVING: REVISITING BULGARIA'S REVENUE FORECASTING MODELS

**Fabio Ashtar Telarico**

Faculty of Social Sciences at the University of Ljubljana (Slovenia)

fabio-ashtar.telarico@fdv.uni-lj.si

0000-0002-8740-7078

*Abstract:* In the thirty years since the end of real socialism, Bulgaria's went from having a rather radically 'different' tax system to adopting flat-rate taxation with marginal tax rates that fell from figures as high as 40% to 10% for both the corporate-income tax and the personal-income tax. Crucially, the econometric forecasting models in use at the Bulgarian Ministry of Finance hinted at an increase in tax revenue compatible with the so-called 'Laffer curve'. Similarly, many economists held the view that revenues would have increased. However, reality fell short of those expectations based on forecasting models and rooted in mainstream economic theory. Thus, this paper asks whether there are better-performing forecasting models for personal- and corporate-income tax-revenues in Bulgaria that are readily implementable and overperform the ones currently in use. After articulating a constructive critique of the current forecasting models, the paper offers readily implementable, transparent alternatives and proves their superiority.

## Introduction

In the thirty years since the end of real socialism, Bulgaria went from having a rather primitive or radically different tax regime to adopting flat-rate taxation in a manner that is manifestly regressive. By 2006, marginal tax rates which used to be as high as 40% flattened to 10% for both the corporate-income tax (CIT) and the personal-income tax (PIT). Academicians and policy advisors armed with imported ideas (and funds, as Dostena Lavergne, 2010 discussed) promoted these reforms as a sure way to foster growth, increase competitiveness and attract foreign capital money (e.g., Ganev, 2016). Eventually, none of these promises was kept (Ninov, 2019). On the contrary, the flat-tax regime accompanied the steadfast deterioration of macro- and socio-economic indicators as gross domestic product (GDP), disposable income, foreign direct investment (FDI), income inequality.

Indeed, the literature has already discussed several aspects of the flat-tax regime and its introduction (e.g., Karagyozova-Markova et al., 2013; Tanchev, 2016; Tanchev & Todorov, 2019). However, not many have highlighted the role that the Bulgarian Ministry of Finance's (MF) forecasting models played in this policy's adoption and persistence. In fact, the official models corroborated the view of those Bulgarian economists who fall into the fallacy of the *Laffer curve* and prognosed increasing revenues under a flat-rate regime (Gălăbov, 2009; Nenovski & Hristov, 2001; Angelov, 2016; Nikolova, 2016). However, reality fell short of those expectations (Figure 1), as it has happened elsewhere after similar reforms (cf. Alvord, 2020). And official models remain severely ineffective even over short-term periods of relative economic stability and despite the absence of major policy change (see Chabin et al. 2020, p. 18-19). Thus, it is high time to shed a light on the failure of these forecasting models rooted in mainstream economic theory.

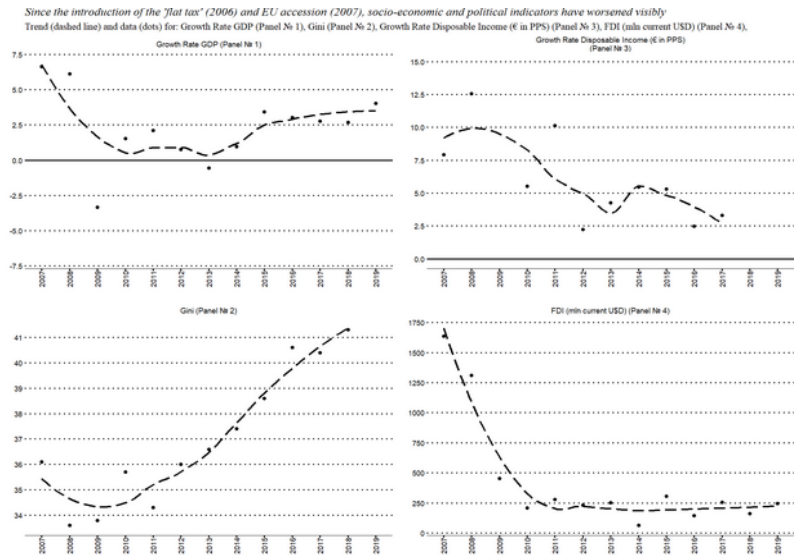

Figure 1 Key socio-economic indicators in Bulgaria, 2007 – 2020.
(Data source by panel: № 1, EUROSTAT, 2021, № 2, 2020; № 3, WB, 2020, № 4, 2021).

# 1. Data and methodology

## 1.1 Data

This paper uses data for both actual and forecasted tax revenues for PIT and CIT for the years 2005–2020. All figures are publicly available in ministerial and parliamentary acts connected with each year's budgetary processes. Intuitively, actual revenues offer a benchmark to assess the efficiency of the both the proposed and the current forecasting models by estimating the appropriate measures of statistical error.

Additionally, several macroeconomic variables are used as proxies representing the entire tax base (the regression models' *predictors*) in the proposed models. Specifically, according to the National Statistical Institute (NSI) just three variables make up over 90% of gross personal income: employment income, pensions (which are tax-exempt), and other social transfers. On the corporate side, the key variables are corporate profits for different categories of companies (non-/financial companies, pension funds, investment firms, insurers) and gross insurance premiums.

## 1.2 Literature review on econometric modelling

According to a literature review commission by the US Federal Reserve (Fukac & Pagan, 2010, p. 2), the '*interpretative* models' that emerge from economic theory are the basis on which forecasting models are built. Each of the former (e.g., Keynesianism, neo-classical synthesis, etc.) roughly corresponds to a 'generation' of the latter. Predictably, electronic calculators' capabilities at a given point in time posited an objective limit to each generation's specific techniques. Hence, it is unsurprising that surveys of the methodological literature agree on which are the main macro-econometric forecasting techniques (Jenkins et al., 2000, pp. 35–47, 48–63, 64–181). Yet, the few endeavours at sketching a typology of these methods lack systematicity (e.g., Bayer, 2013).

Both the current and the proposed models belong to the 'third generation' of forecasting models. Indeed, this class is rather heterogeneous in terms of methods, ranging from differential equations and tax elasticity/buoyancy (for a Bulgarian example see Tanchev & Todorov, 2019) to various autoregressive-moving average (ARMA, ARIMA, ARIMAX) models (on Bulgaria: Telarico, 2021), and many others.

Essentially, the choice of proposing third-generation models is practically and methodologically motivated. On the one hand, it allows a more straightforward comparison with the current ones and makes it easier for forecasting authors to implement them immediately. After all, they provide just a 'small number of simple rules' than can easily be communicated to an auditorium of non-experts (Cairney & Kwiatkowski, 2017, p. 4). On the other, this class of models is preferable to fourth-generation ones from a purely econometric standpoint, too. In fact, comparative analyses and methodological studies found that third-generation models outperform more complex alternatives in terms of sheer efficiency and parsimony (Keene & Thomson, 2007).

### 1.3 An econometric overview of the proposed models

As any 'multiple' or 'multivariable' regression model (MLR), the proposed forecasting models aspire at predicting an independent variable ($y$) by leveraging its relation with some independent variables $(X_1, X_2, \ldots, X_p)$ the value of which is known.[1] Commonly, a MLR model is specified using vectors, as in this paper; but matrix notation is perfectly equipollent. The models are fit to the training data by using ordinary least-square (OLS) regression to estimate the values of the regression coefficients $\left(\beta_{X_1}, \beta_{X_2}, \ldots, \beta_{X_p}\right)$. In addition, the models allow for the dynamic forecasting of revenues thanks to the replacement of the randomly distributed error $\varepsilon$ of OLS regression with the error term $\epsilon$, which is distributed as an autoregressive, moving-average (ARIMA) process. Essentially the introduction of $\epsilon$ allows to incorporate the historical information about the predicted time series ($y_{t'} \forall t' \leq t-1$). Hence the model's equation for each time period is: $y_t = \beta_{X_1} x_{1t} + \beta_{X_2} x_{2t} + \ldots + \beta_{X_p} x_{pt} + \epsilon_t$.

Given that OLS works best in presence of 'stationary' timeseries, several stationarity tests are employed: augmented Dickey-Fuller (ADF), Phillips-Perron (PP), and the Kwiatkowski-Phillips-Schmidt-Shin (KPSS) test. Comparing their results allows to identify distortions due to small sample sizes.

The correction, of trends is operated through differencing (for stochastic trends). In fact, the available literature on Bulgaria suggest that differencing once is usually sufficient (Tanchev, 2016; Tanchev & Todorov, 2019; Telarico, 2021). For determinist trends, natural or base-10 logarithms are not preferable because the transformed series are only covariance-stationary (Kirchgässner et al., 2013, p. 156). Thus, this paper considers also the appropriateness of filters that decompose each timeseries in a trending (non-stationary) component and a stationary one. The literature has made large use of two such tools: the Beverdige-Nelson (BN) filters and the Hodrick-Prescott (HP) filter. Notably, the HP filter has yielded appreciable results in the study of fiscal (Todorov, 2021) and monetary policy (Telarico, 2022) in Bulgaria. However, comparing the two filters' results is indispensable because of the two approaches complimentary weakness: 'artificial short-run cycles due to overdifferencing' (Kirchgässner et al., 2013, p. 161), assumptions forcing the cyclical component's mean to be null, and lack of a unique solution for the BN filter; and incorrect selection of the smoothing parameter's value (Baxter & King, 1999) and endpoints' suboptimality (King & Rebelo, 1993, p. 219) for the HP filter.

Alternatively, this paper acknowledges recent advancements in macro-econometrics arguing that OLS regressions of non-stationary variables are consistent as long as the latter are cointegrated (Kirchgässner et al., 2013, p. 209). Practically, cointegration is assessed quantitatively when there are qualitative reasons to suppose it (e.g., between tax base and tax revenues) by carrying out the eigenvalue version of Johansen's test. Practically, an OLS is attempted for cointegrated, non-stationary variables even if a stationary timeseries can be obtained by differencing or filtering.

### 1.4 Forecasting error

The efficiency of the proposed and current models is assessed using testing data which were not used to train the model and consisting of the last three years of observations. Namely, all the most used forecasting-error's measures are considered (Kirchgässner et al., 2013, pp. 87–88; Chabin et al., 2020, p. 10): the mean error (ME), the sign of which informs about biases; the mean absolute error (MAE), which corrects for error's cancelling-out, but overemphasises underestimations; symmetric absolute errors (sMAE), which correct the MAE's tendency to overweight underestimations; the root mean square error (RMSE), that provides information on the size of the errors on the mean; and Theil's $U_1$, that quantifies the distance between actual and forecasted timeseries.

### 1.5 Specification of the predictors

The predictor variables are selected on both macroeconomics and econometric criteria: (1) macroeconomic relevance, the variables represent a large part of the tax base as defined in the relevant legal act; (2) strong and (3) significant correlation with tax revenues.

The first criterium manifests a macroeconomic rationale rooted in the tax base's higher predictability in comparison to revenues' other determinant. Simplistically, tax revenues under a flat-rate regime depend by the effective tax rate ($r_e$) – which, in turn, is a function of the policy rate ($r_p$) – a and the scope of the tax base ($TB$) according to the formula $T = f(r_e, TB)$. Clearly, the effective and policy rates can shift quite dramatically due to policy changes and is often an unobservable variable dependent on the tax regime's complexity and the tax base's composition. Moreover, there are ongoing policy and theoretical debates regarding the sign of the relation between tax rates and the tax base. Whereas, there is a certain continuity and predictability in the tax base's legal definition due to the need of preserving the tax regime's logicalness and, especially within the EU, ensure international harmonisation (Barrios et al., 2020).

The second criterium provides econometric backing to the previous argument. In fact, statistical indexes of correlation indicate how strong/weak is the relation between the evolution of revenues and of thee selected predictors. In this case, besides Pearson ($r$), two other coefficients are calculated: Kendall ($\tau$), Spearman ($\rho$). Practically, the latter two are better 'at dealing with violations of standard assumptions' (Wilcox et al., 2013, pp. 328–329, 319), with $\tau$ being more robust and slightly more efficient than $\rho$.

The third criterium strengthens the econometric value of the preceding argument by testing the significance of the correlation between predictors and dependent variable. In the text, estimates are for Student's t-test; Pearson's $\chi^2$, the Wilcoxon's signed-rank test, and Mann–Whitney's rank-sum test, for robustness to violations of one of another assumption (cf. Wilcox et al., 2013, p. 323).

## 2. Current forecasting models

### 2.1 Common remarks for PIT and CIT forecasting model

Ultimately, Bulgarian tax forecasts' opacity is one of the explanations for the difficulty in assessing their effectiveness. Moreover, this opaqueness is undeniable considering that the MF did not conduct a methodological review after the Great Recession, even though much more established forecasting institution did so (ARA-PE, 2019; RTMRF Advisory Panel, 2012). For this paper, the Author obtained more detailed information on the current models by filing several requests the disclosure of public information not already published in accordance with Bulgarian law.

As a result, it appears that the forecasting models for PIT and CIT revenues are comprised of two parts: one strictly mathematical; the other somewhat more 'discretionary'. The following paragraphs better detail the implementation of the model and its discretionary adjustments for each tax. Yet, there is a general disclaimer that must be made.

Crucially, despite a direct question, the Ministry's documents fail to spell out the detail of the method used. By reading between the lines of the Ministry's documents, it is reasonable to infer that adjustments related to policy changes are integrated in the models' results only ex-post, (as also Angelov & Bogdanov, 2006, p. 13 supposed) concluded more than a decade ago. Hence, one can infer that the MF's backward forecasting is based on tax-base specific elasticities. Furthermore, the MF's allows significant room for discretionary interventions deriving from enacted and/or planned changes in tax legislation and administrative regulation of the labour market. The only hint as to how these estimates are conducted consists a vague mention of reporting data from the National Revenue Agency [NAP ...] the National Statistical Institute, the Employment Agency, the Bulgarian National Bank [BNB] and other statistical and administrative sources.

### 2.2 Current forecasting model for PIT revenues

As regards PIT, the MF stated in an answer to the Author's requires for the disclosure of public information that its PIT model takes into account 'the relevant indicators from the official macroeconomic forecast of the Ministry of Finance.' Namely, the model considers five independent variables: number of persons employed (EMP); unemployment rate (U); average wage (AWG); compensation of employees (CE); gross domestic product at current prices (BVP). Summarily, the model is based on the sensitivity of PIT revenues to the percentage change in the selected components of its tax base as follows:

$$PIT_t = \eta_{EMP}EMP_t + \eta_U U_t + \eta_{AWG}AWG_t + \eta_{CE}CE_t + \eta_{BVP}BVP_t$$

$$where: \quad \begin{array}{l} PIT_t \text{ are personal income tax renevues at the time } t \\ \eta_X \text{ is the elasticity of renevues to the predictor } X \end{array} \quad (1)$$

Unfortunately, historical estimates of elasticity ($\eta$) are obtained using an undisclosed methodology. So, it is impossible to reproduce official forecasts even if these five components of the PIT's tax base were known.

### 2.3 Current forecasting model for CIT revenues

As regards CIT, the methodology is neither clear nor stable through the years. For instance, in 2021 the MF stated that its econometric model forecasts the three main corporate taxes (KD), the tax on dividends (DDD), and that on insurance premiums (DZP) through a single backward-looking model considering:

> the tax rate, the nominal growth of the gross operating surplus[, ...] the declared taxable profit/loss for the [previous] financial year [...], as well as the data declared by taxable persons with the annual tax returns for losses that are deductible in subsequent reporting periods.
>
> (Reshenie # 963 na Ministerskia savet, 2020, pp. 76–77)

This model looks similar to the one employed in 2019 and 2020 (Reshenie # 928 na Ministerskia savet, 2018, pp. 59–60; Reshenie # 815 na Ministerskia savet, 2019, p. 58). However, in 2018 the description was visibly different (Reshenie # 808 na Ministerskia savet, 2017, p. 86). And it changed again in 2022, albeit the legislation remained almost unvaried (Reshenie # 43 na Ministerskia savet, 2022, p. 77 [e-version: 207]).

Overall, the CIT-revenue forecasting model for 2022 is very similar to the PIT one. However, the tax base is here much more complex as it includes: declared profits ($\pi$); advance payments on the KD in accordance with applicable legislation for companies with over 300,000 leva in yearly turnover (ADV); equalisation contribution paid by sole traders (EQC); amounts refunded due to overpayment in the previous year (REF); and the losses carried forwards for tax purposes (TLS). Thus, the model is based on the sensitivity of PIT revenues to the percentage change in the components of its tax base. Additionally, the DZP and most minor corporate taxes are not estimated directly whilst the DDD is forecasted separately, through simpler first-order autoregressive (AR) models.

$$CIT_t = \eta KD_t + DZP_t \eta + DDD_t$$
$$CIT_t = \psi KD_t + DZP_t = \eta_\pi \pi_t + \eta_{ADV} ADV_t + \eta_{EQC} EQC_t + \eta_{REF} REF_t + \eta_{TLS} TLS_t$$
$$DDD_t = AR \eta \eta = \varphi_0 DDD_{t-1} + \varepsilon_t \qquad (2)$$
$$CIT_t \text{ are corporate income tax renevues at the time } t$$
$$where: \quad \eta \quad \eta_X \text{ is the elasticity of revenues to the predictor } X$$
$$\varepsilon_t \text{ are the errors of the autoregressive model}$$

### *2.4 Pros and cons of the current models*

This clarifying overview provides the basis to argue the model's outdatedness and inadequacy. Schematically, the current models offer (1) some practical advantages, essentially related to the limited need for periodical revision. But they suffer from evident drawbacks related to (2) the selection of variables, (3) excessive arbitrariness and lack of transparency, and (4) their underlying econometric functioning.

#### *2.4.1 Pros — Practical advantages*

The advantages of the current models are mostly related to a certain assessment of their econometric implementation. In fact, without requiring frequent updates, the 'frequently-used method of forecasting revenue by applying an aggregate tax buoyancy to GDP forecasts is usually reasonably reliable' (IMF FAD, 2020, p. 2). In fact, according to some estimates, '90% of the explained forecast error' of CIT and PIT 'can be attributed to wrong macroeconomic assumptions' rather than wrong elasticity estimations (Göttert & Lehmann, 2021, p. 20).

Moreover, assuming that elasticity is a long-run relation, its value is stable unless there is a structural shock (Jenkins et al., 2000, p. 39).

#### *2.4.2 Cons — Variable selection*

As regards variable selection, different problematic aspects emerge for PIT and CIT forecasts. Generally, both models are not transparent enough to allow anyone to reproduce the estimates. Moreover, the choice of the current predictors does not seem econometrically sound.

Essentially, it is difficult to find either a reasonable macroeconomic or econometric explanation for these choices. In macroeconomic terms, it is hard to see why CE and U should be highly determining for PIT revenues. It is not even so useful to use unemployment as proxy for social transfers as the MF argues. In fact, the nominal expenditure in taxable social transfers is directly available in advance and can be forecasted by the National Insurance Institute (NOI). Meanwhile, CE is almost completely irrelevant given that it represents 0.008% of Gross National Income. Additionally, EMP and AWG can be considered duplicates, as both stand as proxies for taxes on salaries and employment relations more generally.

As regards CIT, the selection of variables is neither clear nor stable through the years. Despite the complete lack of clarity in the MF's documents, the KD model looks econometrically quite similar to the PIT one. However, the tax base is here much more complex as it includes between five and seven variables. Again, it is difficult to find a reasonable technical explanation for these choices. In macroeconomic terms, it is hard to see why the model would need to consider so many other variables (ADV, EQC, TLS, ADT) when the KD is a tax on profits. Even assuming that more variables would increase precision, other indicators would be more relevant since they are more closely related to business cycles (e.g., GDP). In addition, it makes little sense to consider the aggregated profits for all companies given the differences in different sectors' performances (IMF FAD, 2020, p. 3) and in the applicable tax regimes.

### 2.4.3 Cons — Econometric weakness

All in all, the current model's underlying predictive power rests on the correct estimation of tax elasticity to model the long-term relation between tax base and revenues. But many scholars have had second thoughts about the use of such technique to forecast tax revenues. Namely, there is a convincing econometric argument based on cases of 'false predictions of the elasticities' in developed countries (Göttert & Lehmann, 2021, p. 20). Others built strong cases noting the difficulty of estimating elasticity correctly (Sen, 2006) or the forecasts' scarce precision (Botrić & Vizek, 2012). Moreover, Bulgarian forecasters seem to treat tax-related elasticities as a structural factor. But this assumption has been disproven time and again (Saez et al., 2009, pp. 43–46).

## 3. The new forecasting model for tax revenues

The methodology disclosed by the MF shows that recommendations to scrape elasticity-based models did not exert any effect. Crucially, 'unrealistic forecasts' play a key role in the spreading anti-Keynesian, trickle-down economics 'on the political left and right' and justifying the adoption of RTRs and 'fiscal profligacy' (Frankel, 2008, p. 13).

Hence, it is opportune to verify whether alternative forecasting models can be more effective that the current ones.

### 3.1 Selection of variables for PIT forecasts

The proposed model improves the MF's choice of proxies for the estimation of the PIT's tax base in both macroeconomic and econometric terms. The MLR's regression coefficients are estimated combing the dataset presented above ($n = 24$) and administrative data for selected explanatory variables with 16 yearly observations ($p = 16$) to date. Namely, the model looks as follows:

$$PIT_t = \beta_{WAGE}WAGE_t + \eta_{SOC}SOC_t + \epsilon_t$$

$$\text{where:} \quad \begin{aligned} & PIT_t \text{ are personal income tax renevues at the time } t \\ & \beta_X \text{ is the regression coefficient of the predictor } X \\ & \epsilon_t \text{ is the error term, distributed as an } ARIMA\,(p,d,q) \text{ process} \end{aligned} \qquad (3)$$

As shown below, the choice of these predictors is econometrically sounds. Namely, all predictors are strongly and significantly correlated with revenues, proving a strong statistical rationale to an economically sensible selection.

### 3.1.1 Representativity of the tax base

First, selecting average employment income (WAGE) and taxable social transfers (SOC, which exclude, notably, pensions) allows to account for over 90% of total (monetary and in-kind) taxable average income. Moreover, these variables are much easier to measure and forecast than non-labour income or financial and other sorts of rents. Unfortunately, the NSI's *Infostat* platform only provides data going back to 2004.

### 3.1.2 Strong and significant correlation with PIT revenues

Besides their macro-economic relevance, these two variables are also statistically corelated with revenues. In fact, the aggregate of the considered variables' correlation with PIT revenues is larger than .95 except Pearson's $r$ and Kendall's $\tau$ for SOC.

These results are also highly significant given that both Student's t and Whitney's $U_1$ paired tests allow to reject the null hypothesis of independence with 99% confidence for all the considered variables (including their aggregate).

## 3.2 Selection of variables for CIT forecasts

The proposed model improves the MF's choice of proxies for the estimation of the CIT's tax base in both macroeconomic and econometric terms. The MLR's regression coefficients are estimated combing all the available data for the period 2006-2017 (training data set), the next three observations are kept as testing data.

### 3.2.1 Representativity of the tax base

The corporate-tax regime in Bulgaria is rather fragmented, with ad-hoc taxes weighing on specific sectors and their activities to different extents. However, the KD on corporations' profits accounted for about 90% of the corporate taxes' revenues in 2002–20. Only DZP are estimated separately, given gross insurance premiums (PRM), using a simple linear regression model. Thus, the KD's tax base, non-financial and financial corporations' profits ($\pi = \pi_{NF} + \pi_F$) may be a good proxy for the CIT's base. Namely, given that there are no data on the value of dividends distributed in Bulgaria, total corporate profits are tested for correlation and causation with both KD revenues and the sum of KD and DDD revenues. Hence, the general model for CIT revenues is:

$$CIT_t = \widehat{\pi}KD_t + DDD_t\widehat{\pi} + DZP_t$$
$$CIT_t = \widehat{\pi}KD_t + DZP_t = \beta_{\pi_{NF}}\pi_{NF_t} + \beta_{\pi_F}\pi_{F_t} + \epsilon_t$$
$$DZP_t = \beta_{PRM}PRM_t + \epsilon_t$$

$$where: \quad \widehat{\pi} \quad \beta_X \text{ is the elasticity of renevues to the predictor } X$$
$$\epsilon_t \text{ is the error term, distributed as an } ARIMA\,\widehat{\pi}p, d, q\widehat{\pi}\text{proces}$$

$CIT_t$ are corporate income tax renevues at the time $t$

(4)

Overall, these variables are not only representative of the corporate taxes' base, but also easier to measure and forecast than the ones currently in use. Unfortunately, however, determining the total amount of corporate profits is made somewhat difficult by lack of clear data.

As regards financial institutions, the NSI provides data for three categories of companies: 'pensions funds', 'investment firms' and 'insurance companies'. Thus, the financial sectors' profits equal the sum of these three sectors' profits $\pi_F = \pi_{Insurance} + \pi_{Investment} + \pi_{Pension}$. Yet, profit figures are available only for the first two.

So, insurers' profits ($\pi_{Insurance}$), are calculated on the basis of the Key Economic Indicators for Insurance Enterprises dataset, by diminishing the turnover (TNR) of gross claims incurred (CLM) and purchases of goods and services (PUR), so that: $\pi_{Insurance} = TNR_{Insurance} - \left(CLM + PUR_{Insurance}\right)$.

Estimating non-financial corporations' profits is somewhat less straightforward. In fact, besides PUR, the available datasets contain two similar variables for expense: 'Remuneration expenses' (RXP) and 'Staff expenses' (SXP). Answering to an industrialist lobby's complain about this confusing distinction, the MF stated that: 'There is no legal definition of "staff expenses" in any law, including tax and accounting legislation.' (Karayvanova, 2016, p. 1) Rather, it is the equivalent of the employees' 'Staff income' or the 'Employees' benefits' in the employers' accounts (MS-RB, 2005, pp. 76–84). So, RXP are a wider concept than SXP. Yet, given the special tax treatment reserved to 'compulsory social security contributions' in Bulgaria (see DOPK, 2005/2021 and; ZKBO, 2006/2022 art. 41.3,5–8), it is not possible to choose between the two a-priori. Hence, the following calculations test two definition of non-financial firms' profits and, coherently, profits overall:

$$\pi_{NF1} = TNR_{NF} - ꞮRXP_{NF} + PUR_{NF}Ɲ \;\; \Rightarrow \;\; \pi_1 = \pi_F + \pi_{NF1}$$
$$\pi_{NF2} = TNR_{NF} - ꞮSXP_{NF} + PUR_{NF}Ɲ \;\; \Rightarrow \;\; \pi_2 = \pi_F + \pi_{NF2}$$
$$(5)$$

As shown below, the choice of these determinants is econometrically sounds as there is strong and significant correlation between predictors and revenues.

### 3.2.2 Strong and significant correlation with CIT revenues

Besides their macro-economic relevance, the selected variables are also statistically related to total revenues. First, it is the first definition of non-financial corporation's profits ($\pi_{NF1}$, which uses RXP in the cost component) is slightly more strongly correlated to KD and KD+DDD revenues than the alternative formulation ($\pi_{NF2}$, which uses SXP). Importantly, the considered variables' correlation with the corresponding tax's revenues is very strong no matter what measure is used (on average: $r = 0.913$; $\tau = 0.785$, $\rho = 0.879$). Moreover, there is little difference between the correlation indexes for KD and KD+DDD across tax bases (on average: $\Delta r_\mu = 2.1\%$; $\Delta \tau_\mu = 1.6\%$, $\Delta \rho_\mu = 0.45\%$).

These results are also highly significant given that both Student's t and Whitney's U paired tests allow to reject independence with 99% confidence except $\pi_2 \therefore KD$ ($t = 0.124 ; U = 0.01245$).

### 3.3 PIT model estimation

### 3.3.1 Multivariate linear regression

The ADF, PP, and KPSS tests' discordant results impede to identify a set stationary timeseries for OLS, hinting at both HP-filtered and first-difference ($I(1)$) data. Moreover, Johansen's test shows significant cointegration at level ($I(0)$). Thus, OLS is run for each as shown in Table 1.[2]

Table 1 Estimations of the proposed PIT models.

| Variables | log likelihood | AIC | AICc | BIC | $R^2$ |
|---|---|---|---|---|---|
| I(1) | -85.38 | 176.7 | 179.42 | 178.4 | .01 |

| | | 6 | | 5 | |
|---|---|---|---|---|---|
| HP | -51.81 | 117.61 | 136.28 | 122.08 | .99 |
| I(0) | -91.90 | 189.79 | 192.19 | 191.71 | .93 |

Clearly, the OLS MLR using HP-filtered variables is the most effective. Its $R^2$ is very close to the unit: approximately 99% of the resulting variability is explained by the proposed model. Moreover, it also has the largest log-likelihood and the most favourable (smallest) information criteria.

### 3.3.2 PIT revenue forecasting error

Using the testing dataset for WAGE and SOC, the proposed PIT model appears superior to the current one regardless of the metrics (Table 2). In particular, despite training on HP-filtered data, the proposed model is more efficient than the current one also when the results are compared with at-level revenues.

Table 2 Forecasting errors for the current and proposed models (HP-filtered).

| Error: | ME | MAE | sMAE | RMSE | $U_1$ |
|---|---|---|---|---|---|
| Actual Forecasts | 98.18 | 204.69 | 0.05 | 214.55 | 0.03 |
| Proposed model | 89.67 | 89.67 | 0.02 | 95.87 | 0.01 |
| Accuracy gain | 8.51 | 115.02 | 0.03 | 118.68 | 0.02 |

Clearly, the proposed model is more efficient in predicting the mean (much smaller RMSE) and virtually perfect on average even though overestimations do not cancel out underestimations as often (smaller MAE). As suggested by the smaller $U_1$, the proposed model would have overestimated yearly revenues by an average 3.55mln in 2017–20 with a 67% relative increase in efficiency.

### 3.4 CIT model estimation

#### 3.4.1 Multivariate linear regression

The ADF, PP, KPSS, and DF-GLS tests' discordant results impede to identify a set stationary timeseries for the dependent variables KD+¿DDD, DZP, hinting at I(1) and HP-filtered data for the former ($\pi_{NF1}, \pi_F$) and the natural logarithm for the latter. ( $PRM$ ). However, Johansen's test shows significant cointegration for both sets of predictors at level. Thus, OLS is run for these combinations.

Table 3 Estimations of the proposed KD+DDD model at I(0), I(1), and HP filtered.

| Variables | log likelihood | AIC | AICc | BIC | $R^2$ |
|---|---|---|---|---|---|
| I(1) | -64.38 | 134.76 | 137.16 | 136.67 | .70 |
| HP | -59.41 | 130.81 | 141.31 | 135.06 | .95 |
| I(0) | -38.06 | 82.13 | 84.79 | 83.82 | .80 |

Clearly, the OLS MLR using HP-filtered variables is the most effective. Both its $R^2$ and all the information criteria favour it over the alternatives.

Coming to DZP revenues, the model the natural logarithm of the predictors is decidedly effective, with $R^2 = .9$, the favour of all information criteria, and the highest log likelihood.

Table 4 Estimations of the proposed DZP model at I(0) and after natural logarithm

| Variables | log likelihood | AIC | AICc | BIC | $R^2$ |
|---|---|---|---|---|---|
| I(1) | -13.81 | 31.62 | 34.62 | 31.51 | 0.86 |
| ln | 8.83 | -11.67 | -3.67 | -11.83 | 0.90 |

### 3.4.2 CIT revenue forecasting error

Using the testing dataset for $\pi_{NF1}$ and $\pi_F$ built as mentioned above, the proposed model appears superior to the current one regardless in forecasting KD and DDD revenues. In particular, despite training on HP-filtered data, the proposed model is more efficient than the current one also when the results are compared with at-level revenues (Table 5).

Table 5 Forecasting errors for the current and proposed KD+DDD models (HP-filtered).

| Error: | ME | MAE | sMAE | RMSE | $U_1$ |
|---|---|---|---|---|---|

| Actual Forecasts | 6.53 | 73.33 | 0.03 | 82.69 | 0.02 |
| Proposed model | 5.66 | 69.66 | 0.02 | 76.52 | 0.01 |
| Accuracy gain | 0.87 | 3.67 | 0.01 | 10.17 | 0.01 |

Clearly, the proposed model is more efficient in predicting the mean (smaller RMSE), albeit similarly imprecise (comparable ME) and slightly more biased upwards (slightly smaller MAE). Yet, the proposed model is relatively more efficient overall (smaller $U_1$).

Table 6 Forecasting errors for the current and proposed DZP models compared to at-level data.

| Error: | ME | MAE | sMAE | RMSE | $U_1$ |
|---|---|---|---|---|---|
| Actual Forecasts | 39.47 | 39.47 | 1.69 | 39.66 | 0.85 |
| Proposed model | 36.63 | 36.63 | 0.07 | 3.24 | 0.04 |
| Accuracy gain | 2.83 | 2.83 | 1.62 | 36.41 | 0.81 |

Coming to the simple linear regression model for DZP revenues, the proposed model is clearly more efficient. Namely, it approximates the mean better (smaller RMSE) and only has a large ME because its errors do not cancel each other out as often (smaller MAE and sMAE). Finally, Theil's $U_1$ shows is an appreciable improvement in accuracy — albeit not a massive one.

### 3.5 Reasons to adopt the proposed models

Schematically, the proposed models offer (1) significant practical advantages related to the transparent and simpler underlying econometric functioning; (2) a better selection of variables; (3) increased precisions; (4) adaptability to different tax regimes. Still, they would not necessarily removing all arbitrary adjustments, leaving some degree of flexibility. Yet, (5) the authorities would need to forecast variables they have little experience with.

*3.5.1 Econometric simplicity and transparency*

The main econometric strength of the proposed models lies in their being extremely transparent, as the MF would not even need to publish the regression coefficients estimated for the MLR. Instead, publishing the predictors' forecasts would allow anyone who has access to statistical tools (e.g., R, Stata or even Excel) to verify their correctness. This is essential in today's political climate because, transparency is antagonistic to policy-based modelling as transparent models cannot be used to preserve and rationalise biased, partisan opinions by cherry-picking favoured results.

In addition, the proposed models are quite simple, which is often associated with efficiency and parsimony, whereas 'overelaboration and overparameterization is often the mark of mediocrity.' (Box, 1976, p. 796).

### 3.5.2 Improved selection of variables

The proposed model improves the MF's choice of predictors for the estimation PIT and CIT revenues base in both macroeconomic and econometric terms.

In fact, the variables chosen to forecast PIT revenues are much easier to measure and forecast than non-labour income or financial and other sorts of rents. Additionally, this choice forces to diversify data source, as the data is forecasted separately by the MF (WAGE) and the NOI (SOC), while the NSI operates an ex-post revision. Thus, the proposed model would increase public scrutiny. In addition, these variables are also statistically correlated to total revenues. The relation is very strong and highly significant no matter what measure is used.

### 3.5.3 Increased precision

The proposed model would have realised a 67% relative increase in precision. Namely, the proposed PIT model is more efficient in predicting mean revenues over a period of time (smaller RMSE) and virtually perfect on average ($ME \approx 0$) even though overestimations do not cancel out underestimations as often (smaller MAE). Meanwhile, the proposed CIT model is more efficient in predicting the mean (smaller RMSE), albeit similarly imprecise on the long term (comparable ME) because overestimations do not cancel out underestimations as often (slightly smaller MAE). Yet, there is a sensible increase in relative efficiency (smaller $U_1$), suggesting that simplicity and transparency are not alternative, but complementary to precision.

## Conclusion

Building on previous, more limited studies, this paper showed that using complex, opaque and inefficient forecasting models for forecasting revenues is the prevailing economic practice in official Bulgarian forecasts.

Practically, this paper offers the first systematic overview and attempt at formalising the current forecasting models used by the Bulgarian Ministry of Finance. In addition, it provides three alternative MLR models to forecast PIT, KD+DDD and DZP revenues in a transparent and parsimonious way using a limited number of variables as proxies for the respective tax bases. Crucially, this class of models are comparatively easier to implement than the one already provided in the literature (e.g., Telarico, 2021's ARIMA model). Using established measures of forecasting error and testing data, the proposed models are show to be more efficient than the current ones in forecasting revenues under many regards.

In terms of limitations, it is worth underlining that this paper does not argue that MLR models are the *most* efficient for forecasting revenues in Bulgaria. But just that they are *more* efficient that the current ones. Thus, more studies are needed to find out how different techniques (e.g., non/Bayesian VAR, pure ARIMA, generalised linear regression, etc.) as well as fourth-generation dynamic stochastic general equilibrium compare to the proposed MLR. Contextually, arguments can be made as to the best tools to measure the model's precision and econometric soundness (e.g., errors measures, the Granger test, or the Toda-Yamamoto test, etc.). However, something emerges as strikingly undeniable. In fact, despite the absence of massive policy changes and exogenous shocks, the current models are less effective than simpler alternatives using easily forecastable predictors. Thus, it seems apparent that the bulk of the current forecasts' imprecision stems from a wrong design, rather than substantial unpredictability.

Hence, the findings presented above add weight to respected Bulgarian economists' remarks (e.g., Angelov & Bogdanov, 2006; Gechev, 2010; Tanchev, 2016) that encourage the MF to update its methodologies. Undoubtedly, the economic profession will have a role to play in discontinuing the current models. However, doing so will require an improved understanding of the unspoken assumptions and the implicit ideology that underpins (see Nikolov, 2008, p. 100ff) the current models. Hence, it is necessary to underline the role that these models have had in supporting 'irresponsible, socially unjust and technically ineffective' (Nikolov, 2008, p. 99) economic policies in Bulgaria. After all, unrealistic forecasts have played a key role in justifying the adoption of fiscal profligacy elsewhere in the past. And forecasting models similar to the ones currently in use in Bulgaria 'have been used by various countries' to justify regressive tax regimes (Jenkins et al., 2000, p. 40). Hence, they play a crucial role in spreading trickle-down economics across the political spectrum (Frankel, 2008, p. 13) and in academia. For better or worse, much is still to be written on this.

## Endnotes

[1] One must notice that sometimes MLR is improperly labelled *multivariate* regression by behavioural and social scientist, making the words almost synonyms (cf. Arminger et al., 1995, pp. 97–99; Charles, 2012, fol. 1,4 for examples of such an improper use).

[2] AIC, AICc, and BIC are information criteria (Akaike's, Akaike's corrected for small samples, and Bayesian) that quantify two or more models' efficiency in relative terms.

## Conflict of Interest

Nothing to declare.